\documentclass[aip,apl,superscriptaddress,reprint,twocolumn,longbibliography,amsmath,amssymb]{revtex4-1}
\usepackage{graphicx}  
\usepackage[toc,page]{appendix}
\usepackage{epstopdf}
\usepackage{dcolumn}   
\usepackage{bm}        
\usepackage{amsmath}
\usepackage{verbatim}   
\usepackage{setspace}

\begin{document}

\title{Thickness dependence of unidirectional spin-Hall magnetoresistance in metallic bilayers}
\author{Yuxiang Yin}
\email[E-mail: ]{y.yin@tue.nl}
\author{Dong-Soo Han}
\author{Mark C. H.  de Jong}
\author{Reinoud Lavrijsen}
\affiliation{Department of Applied Physics, Eindhoven University of Technology, PO Box 513, 5600 MB Eindhoven, The Netherlands}
\author{Rembert A. Duine}
\affiliation{Department of Applied Physics, Eindhoven University of Technology, PO Box 513, 5600 MB Eindhoven, The Netherlands}

\affiliation{Institute for Theoretical Physics and Center for Extreme Matter and Emergent Phenomena,
Utrecht University, Leuvenlaan 4, 3584 CE Utrecht, The Netherlands}
\author{Henk J. M. Swagten}
\author{Bert Koopmans}
\affiliation{Department of Applied Physics, Eindhoven University of Technology, PO Box 513, 5600 MB Eindhoven, The Netherlands}

\begin{abstract}
A nonlinear magnetoresistance - called unidirectional spin-Hall magnetoresistance - is recently experimentally discovered in metallic bilayers consisting of a heavy metal and a ferromagnetic metal. To study  the  fundamental mechanism of the USMR,  both  ferromagnetic and heavy metallic layer thickness dependence of  the USMR  are  presented  in  a  Pt/Co/AlOx  trilayer  at  room  temperature. To avoid ambiguities, second harmonic Hall measurements are used for separating spin-Hall  and  thermal contributions to the non-linear magnetoresistance. The experimental  results are fitted  by  using  a  drift-diffusion theory, with  parameters  extracted  from  an analysis  of  longitudinal  resistivity  of  the  Co  layer  within  the  framework  of  the  Fuchs-Sondheimer model. A good  agreement with  the  theory  is  found,  demonstrating  that  the  USMR  is  governed  by  both the spin-Hall effect in the heavy metallic layer and the metallic diffusion process in the ferromagnetic layer. 
 \end{abstract}

\maketitle
In the field of Spintronics, a new way of spin control based on the spin-Hall effect recently has attracted a great deal of attention. It originates from the spin-orbit (SO) interaction which converts a charge current into a net flow of spin angular momentum, exerting a SO torque on the magnetization. This leads to an energy-efficient way of writing information to magnetic memories by switching a magnetic entity via sending a charge current through a nearby nonmagnetic metal\cite{miron2011perpendicular,Liu2012}. Apart from writing, a possible way of reading the memory could be achieved by measuring a so-called spin-Hall magnetoresistance (SMR)\cite{Kobs2011,Nakayama2013}, i.e., a change in electrical resistance due to the spin Hall effect when the spin-accumulation is perpendicular or parallel to the magnetization. Although SMR provides a promising way towards reading memory devices using a two-terminal architecture, the fact that it can only distinguish between the perpendicular and parallel states limits its application.

Very recently, a unidirectional contribution to magnetoresistance - called unidirectional spin-Hall magnetoresistance (USMR) - has been reported in a ferromagnetic/heavy metallic (FM/HM) bilayer structure\cite{Avci2015,olejnik2015electrical,yasuda2016large}. Being different from the ordinary SMR, the resistance changes by reversing the magnetization or the current direction, which could be potentially utilized for reading operation. Based on a drift-diffusion-relaxation theory\cite{Zhang2016a}, this nonlinear behavior is attributed to the dependence of electron mobility on spin-polarization, which is tuned by the spin-Hall effect induced spin accumulation. This spin accumulation is limited to a thin region at the FM/HM interface due to a finite spin diffusion length in both layers, leading to a non-trivial FM and HM thickness dependence of the USMR. So far, this particular dependence on thickness is not evidenced by any experiments. Thus, a systematic investigation of how USMR depends on the layer thickness is urgently needed not only for a better understanding of the origin of USMR, but also for the enhancement of USMR in practical applications.

In this paper, we present the FM and HM layer thickness dependence of USMR in Pt/Co/AlOx trilayers at room temperature. The experimental results are fitted by using the aforementioned drift-diffusion-relaxation theory, with
parameters extracted from an analysis of the longitudinal resistivity of the Co layer within the framework of the Fuchs-Sondheimer model\cite{Kobs2016}. Furthermore, second harmonic Hall measurements enable us to disentangle spin-Hall and thermal gradient contributions to the non-linear magnetoresistance, allowing for a more precise fitting. Good agreement with the theory is found, demonstrating that the USMR depends on both the spin-Hall effect in the HM layer and the electron spin diffusion and relaxation in the FM layer.

\begin{figure}
\centering
\includegraphics[scale=0.31]{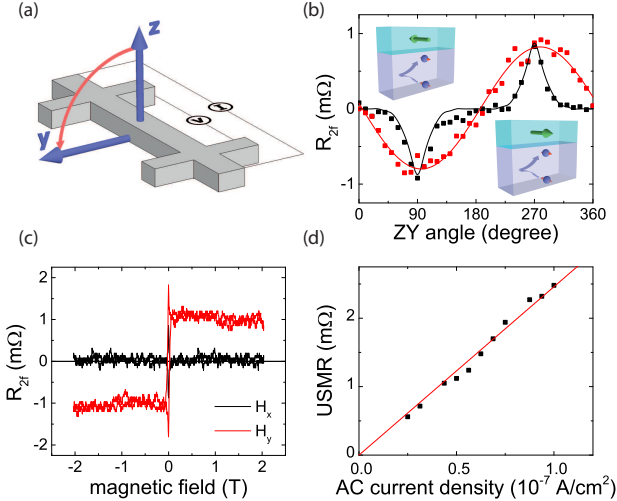}
\caption{\label{Figure1} (a) A schematic illustration of the Hall bar used in the experiment. The measurement scheme of longitudinal resistances is shown. (b) Angle dependence measurements of $R_{2f}$ in the sample Pt(4 nm)/Co(1 nm)/AlOx(1.15 nm) (black dots) and Pt(4 nm)/Co(2 nm)/AlOx(1.15 nm) (red dots). Solid lines represent the fitting results. (c) $y$ (red line) and $x$ (black line) field dependence measurements of $R_{2f}$ in the sample Pt(4 nm)/Co(2 nm)/AlOx(1.15 nm). (d) USMR as a function of current density.}
\end{figure}

For the measurement of magnetoresistance, the multilayer structures are Pt(1-8 nm)/Co(4 nm)/AlOx(1.15 nm) (6 samples) and Pt(4 nm)/Co(1-50 nm)/AlOx(1.15 nm) (11 samples), where we either vary the Pt or the Co thickness as indicated by the thickness range in the parenthesis. These multilayer are then patterned in a form of a Hall bar, shown in Fig. \ref{Figure1}(a), by using electron-beam lithography and lift-off. The length of the Hall bar is 100 $\rm{\mu m}$, the lateral width 5 $\rm{\mu m}$ and the spacing between two Hall bars 20 $\rm{\mu m}$. The samples are deposited on Si/SiO$_2$ substrates by DC magnetron sputtering. Pt was deposited at a rate of 0.08 nm/s, Co was sputtered at a rate of 0.05 nm/s. After deposition, a 1.15 nm thick Al capping layer was finally deposited and further oxidized (by using plasma oxidation during 90 sec at $1\times10^{−1}$ mbar) on top of the Pt/Co stack, to prevent oxidation of the Co layer in air.

The magnetoresistance measurements presented in this work were performed at room temperature by using an AC current source with a current density of $1\times10^7$ A/cm$^2$ modulated at $f$ = 801 Hz. The second harmonics component of the longitudinal resistance $R_{2f}$ is recorded during sample rotation or sweeps of an external magnetic field.

For the measurement of USMR, first an angle dependence measurement of $R_{2f}$ is performed. An external magnetic field $B_{\rm ext}=2$ T, which is high enough to saturate the magnetization in every configuration, is rotating in the $yz$ plane while $R_{2f}$ being recorded at the same time. The results for two types of samples are plotted in Fig. \ref{Figure1}(b). The black dots show the results for a Pt(4nm)/Co(1nm)/AlOx(1.15nm) stack with perpendicular anisotropy (PMA) while the red dots show the results for a Pt(4nm)/Co(2nm)/AlOx(1.15nm) stack with in-plane anisotropy. Starting from the $+z$ direction, $R_{2f}$ gradually decreases and reaches a minimum when the field is along the $+y$ direction. After that, $R_{2f}$ starts to rise and reaches a peak at $-y$. Finally, it returns to its original value after a full rotation. The measurement evidences a resistance contribution that depends on the sign of $M_y$, and the difference of $R_{2f}$ between the $+y$ and $-y$ direction is defined as the USMR. This is further confirmed by a good agreement between the data and a fitted line with respect to $M_y$ (obtained by an anomalous Hall effect measurement), plotted as a solid line in Fig. \ref{Figure1}(b). Note that the transition for the PMA stack is significantly sharper than the in-plane sample, since a high field is needed to pull the magnetization in the plane for a stack with PMA.

To further investigate the USMR, we have measured $R_{2f}$ while sweeping the external magnetic field along transverse ($y$) and longitudinal ($x$) direction. Fig. \ref{Figure1}(c) shows that $R_{2f}$ is constant as a function of $y$ field and reverses sign upon sweeping the field from $y$ direction to $-y$ direction. Two spikes are observed near zero field due to the formation of magnetic domains during magnetization switching. In contrast to a field in the $y$ direction, no difference is measured between the $x$ and $-x$ direction, indicating that, as expected, the USMR only exists in the transverse direction. Compared with the angle dependent measurement, where one have to ensure that the field is strong enough to saturate the sample in the $z$ direction, the field dependent measurement serves as an more efficient way of quantifying the USMR. Thus, in the following, USMR will be obtained by sweeping the field. As a further test, Fig. \ref{Figure1}(d) shows the current dependence of USMR measured in this way, which is linear with the injected current density and converges to zero for decreasing current, since the spin accumulation at the interface scales with the current density.

\begin{figure}[htb]
\centering
\includegraphics[scale=0.45]{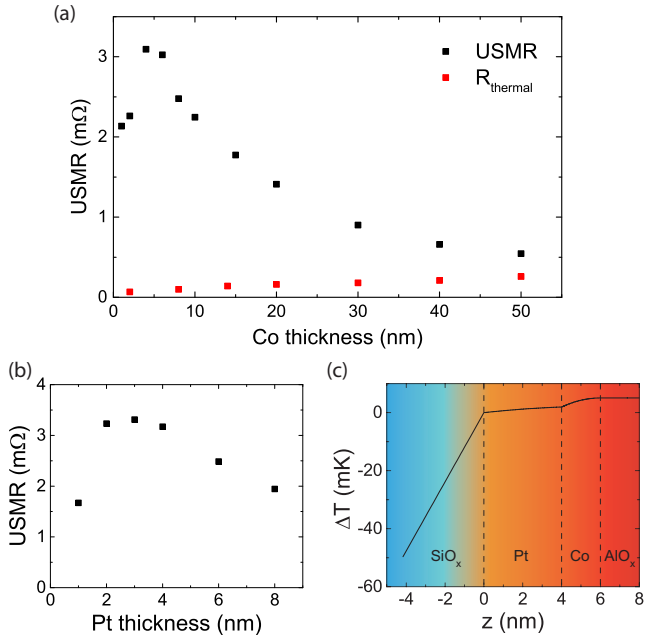}
\caption{\label{Figure2} (a) USMR as a function of the Co thickness and a resistance contribution originated from the thermal effect for Pt(4 nm)/Co(1-50 nm)/AlOx(1.15 nm). (b) USMR as a function of the Pt thickness for Pt(1-8 nm)/Co(4 nm)/AlOx(1.15 nm). (c) Temperature profile in the sample Pt(4 nm)/Co(2 nm)/AlOx(1.15 nm) simulated by using finite element method.}
\end{figure}

To verify the role of the interfacial spin accumulation due to the SHE we examined the dependence of the USMR on the thickness of the NM and FM layers. Fig. \ref{Figure2}(a) and (b) show the absolute change of second harmonic resistance ${\rm USMR} = R_{2f}(+{\rm M})-R_{2f}(-{\rm M})$ measured at constant current density as a function of the Co and Pt thickness. Both curves exhibit qualitatively similar behavior: an initial sharp increase and a gradual decrease as the layer becomes thicker. Apart from the USMR, thermal effects could also contribute to the $R_{2f}$. Thus, exclusion of these thermal effects are required before further analysis, which will be discussed in the following part.

Fig. \ref{Figure2}(c) shows the temperature profile $T(z)$ in a line along the thickness direction of the nanowire by using simulation software suites Comsol multiphysics 5.2. The maximum temperature is found at the top owing to the fact the heat dissipation is faster through the bottom substrate than through the top ambient air. A temperature gradient in the $z$ direction will generate an electric current in the same direction, which interacts with the FM layer through the anomalous Hall effect and generates an electric field ($\propto\Delta T \times M_{\rm sat}$) in $x$ direction\cite{nernst1887electromotorischen}. This will cause a resistance change in $x$ direction and possesses the same symmetry as that of the USMR. In order to separate the thermal contribution from the USMR, we measure the second harmonic Hall resistance to quantify the thermal resistance\cite{Avci2014a} (see Supplementary Material for details). As plotted in the same figure of USMR, see the red dots in Fig. \ref{Figure2}(a), the thermal contributions are found to be increased with the thickness. The maximal thermal resistance is observed at Co thickness of 50 nm, which accounts for about 50\% of the USMR and the ratio is smaller for thinner Co. In following part, the thermal resistance will be subtracted from the USMR to achieve an accurate analysis. 

\begin{figure}[tb]
\centering
\includegraphics[scale=0.37]{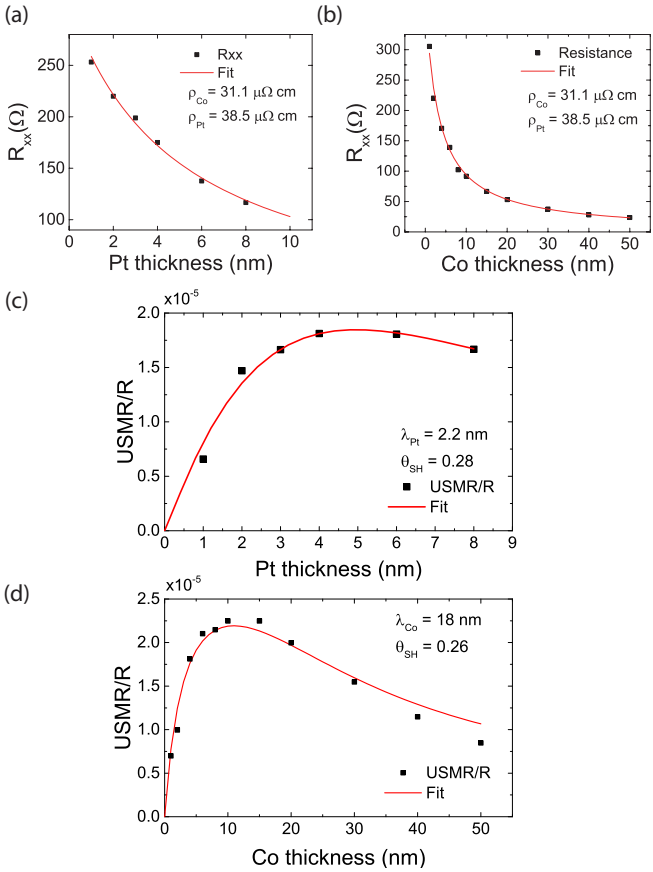}
\caption{\label{Figure3} (a)(b) Longitudinal resistance as a function of the Co and Pt thickness together with a theoretical fitting curve. (c)(d) Normalized USMR as a function of the Co and Pt thickness together with a theoretical fitting curve.}
\end{figure}

In order to compare the experimental measurement of USMR with the model, we first convert the absolute USMR into normalized USMR, i.e. USMR divided by normal longitudinal resistance. For this purpose, the longitudinal resistance $R_{xx}$ is measured and plotted versus Pt and Co thickness in Fig. \ref{Figure3}(a)(b). The plot reveals that the resistance monotonically decreases with thicknesses. The solid line represents the fit which utilizes the Fuchs-Sondheimer approach to extend the conventional $t^{-1}$ resistance model by considering the scattering at the two Co/Pt interfaces\cite{Kobs2016}. The fit describes the experimental data well and gives the bulk resistivity $\rho_{\rm Pt}$ = 37.5 $\mu\Omega$ cm and $\rho_{\rm Co}$ = 31.1 $\mu\Omega$ cm, which are comparable to the values in the literature  \cite{agustsson2008electrical,carcia1988perpendicular}.

Next, we examine the dependence of USMR on the Pt and Co thickness in Pt/Co/AlOx samples. As shown in Fig. \ref{Figure3}(c), the normalized USMR is the largest for a Pt thickness of about 5nm and is reduced for a thicker or thinner Pt layer. Its strong thickness dependence shows that USMR in the structures is mainly influenced by the SHE in Pt. USMR decreases when the Pt layer is thinner than the spin diffusion length due to the reduced spin current caused by back refection at the interface. On the other hand, for a thicker Pt layer, USMR is also reduced by current shunting. A similar behavior is found for USMR upon varying the Co thickness, as shown in Fig. \ref{Figure3}(d), although the maximal USMR is now reached at a Co thickness of 10 nm. Above all, the qualitative behavior matches the prediction of the drift-diffusion-relaxation model. In addition, we  also make a quantitative comparison of the experimentally observed values with the model, which describes the USMR as\cite{Zhang2016a}:
\begin{widetext}
\begin{equation}
\frac{\rho(E)-\rho(-E)}{\rho(E)}= \frac{6 \theta L_F L_H ({p_\sigma - p_N }) {\sigma_{\rm F}} \epsilon  \tanh \left(\frac{d_{\rm F}}{L_{\rm F}}\right) \tanh \left(\frac{d_{\rm H}}{2 L_{\rm H}}\right)}{{\epsilon_{\rm F}} ({d_{\rm F}} {\sigma_{\rm F}}+{d_{\rm H}} {\sigma_{\rm H}}) \left(\frac{{L_{\rm H}} \left(1-{p_\sigma }^2\right) {\sigma_{\rm F}} \tanh \left(\frac{{d_{\rm F}}}{{L_{\rm F}}}\right) \coth \left(\frac{{d_{\rm H}}}{{L_{\rm H}}}\right)}{{L_{\rm F}} {\sigma_{\rm H}}}+1\right)},
\label{Equation USMR}
\vspace{1em}
\end{equation}
\end{widetext}
\noindent where $d_F$ ($d_H$) is the thickness of the FM (HM), $L_F$ ($L_H$) is the spin diffusion length of the FM (HM), $\sigma_F$ ($\sigma_H$) is the conductivity of the FM (HM), $\theta$ is the spin Hall angle of the HM, $\epsilon$ is the electric field in FM, $\epsilon_F$ is the Fermi energy, $p_\sigma$ is the conductivity spin asymmetry, and $p_N$ is the difference of density of states at Fermi energy. In our sample, the Fermi energy $\epsilon_F=5$ eV and spin asymmetry $p_\sigma-p_N=0.5$\cite{Zhang2016a}.

\vspace{1cm}

By fitting the thickness dependence of the normalized USMR to Eq. \ref{Equation USMR} (red line in Fig. \ref{Figure3}), it can be seen that the specific behavior of the data is in line with the drift-diffusion-relaxation model based on the spin-Hall effect, and a spin Hall angle of 0.3 for Pt, and a spin diffusion length of 18 nm and 2.2 nm for Co and Pt, respectively, can be extracted. The spin diffusion lengths are similar to the literature values\cite{zhang2013determination,piraux1998temperature}. This implies that the USMR in the Pt/Co system is governed by 1) spin-Hall effect in the Pt layer and 2) electron spin diffusion and relaxation in the Co layer. In a recent work\cite{kim2016current} which uses a similar structure (Py/Co), a unidirectional contribution was found in the first harmonic resistance by applying a high current density ($J\sim 10^8$ A/cm$^2$). In this experiment, magnon excitation, instead of an electronic diffusion-relaxation process, is claimed to attribute to this first harmonic USMR. It is also found that the magnon-induced USMR increases with increasing temperature. We do not intend to conduct a temperature dependence measurement here, due to the complex variation of all transport parameters with temperature \cite{isasa2015temperature,ueda2012temperature,Kim2015} (polarization, spin diffusion length, spin Hall angle and conductivity) in the drift-diffusion-relaxation model (Eq. \ref{Equation USMR}), which will make it extremely difficult to draw pertinent conclusions. Moreover, in the magnon experiment\cite{kim2016current}, a much higher current density is used compared to ours, and only addressed the first harmonic in the resistance, which further complicates a meaningful comparison.

To disentangle electronic and magnonic contribution, a measurement of the temperature dependent of the USMR needs to be performed, which is beyond the scope of this paper. For the electron contribution described before (Eq. \ref{Equation USMR}), taking into account the temperature variation of all transport parameters (polarization, spin diffusion length, spin Hall angle and conductivity) predicts that the USMR decreases with increasing temperature \cite{isasa2015temperature,ueda2012temperature,Kim2015}, whereas the magnon-induced USMR behaves oppositely\cite{kim2016current}.

The extracted room-temperature spin Hall angle in Pt appears to be higher than the value $\sim$0.1 measured in other work\cite{Liu2012a,zhang2015role}. We still think this model captures the essential physics of the observed effect, although full quantitative agreement cannot be reached due to various reasons. One reason is the simplifications of the model by assuming a spherical Fermi surfaces and constant density of states at the Fermi energy\cite{Zhang2016a}, which underestimates the magnitude of USMR. Moreover, the model\cite{Zhang2016a} includes only spin-dependent scattering in the bulk of the ferromagnetic layer. Like in the giant magnetoresistance effect, however, both bulk and interface scattering can contribute to the USMR \cite{Avci2015}. The underestimation would be more if the spin-mixing conductance is incorporated, since the Pt/Co interface is regarded as fully transparent in the model, i.e., the spin-mixing conductance is infinite. Finally, additional charge-spin conversion can take place at either the Pt/Co or Co/AlOx interface, that may lead to a larger spin-Hall effect \cite{amin2016spin,wang2016giant}.

In conclusion, USMR is observed in Pt/Co/AlOx systems and we have shown that the dependence of the USMR on the thickness of both the HM and FM layers agree qualitatively with the theory based on electron spin drift-diffusion-relaxation model. We believe this result provides a better understanding of the physical origin of the USMR and is of importance for its possible applications in spintronic devices.

\section*{Supplementary Material}
See Supplementary Material for the quantification of thermal contributions to the unidirectional spin-Hall magnetoresistance by measuring the second harmonic Hall resistance.
\section*{Acknowledgement}
This work is part of the research program electrical control of spin dynamics and nanomagnetic devices with project number FOM-11PR2906, which is (partly) financed by the Netherlands Organisation for Scientific Research (NWO).
\bibliographystyle{apsrev4-1}
\bibliography{USMR}

\begin{thebibliography}{23}%
\makeatletter
\providecommand \@ifxundefined [1]{%
 \@ifx{#1\undefined}
}%
\providecommand \@ifnum [1]{%
 \ifnum #1\expandafter \@firstoftwo
 \else \expandafter \@secondoftwo
 \fi
}%
\providecommand \@ifx [1]{%
 \ifx #1\expandafter \@firstoftwo
 \else \expandafter \@secondoftwo
 \fi
}%
\providecommand \natexlab [1]{#1}%
\providecommand \enquote  [1]{``#1''}%
\providecommand \bibnamefont  [1]{#1}%
\providecommand \bibfnamefont [1]{#1}%
\providecommand \citenamefont [1]{#1}%
\providecommand \href@noop [0]{\@secondoftwo}%
\providecommand \href [0]{\begingroup \@sanitize@url \@href}%
\providecommand \@href[1]{\@@startlink{#1}\@@href}%
\providecommand \@@href[1]{\endgroup#1\@@endlink}%
\providecommand \@sanitize@url [0]{\catcode `\\12\catcode `\$12\catcode
  `\&12\catcode `\#12\catcode `\^12\catcode `\_12\catcode `\%12\relax}%
\providecommand \@@startlink[1]{}%
\providecommand \@@endlink[0]{}%
\providecommand \url  [0]{\begingroup\@sanitize@url \@url }%
\providecommand \@url [1]{\endgroup\@href {#1}{\urlprefix }}%
\providecommand \urlprefix  [0]{URL }%
\providecommand \Eprint [0]{\href }%
\providecommand \doibase [0]{http://dx.doi.org/}%
\providecommand \selectlanguage [0]{\@gobble}%
\providecommand \bibinfo  [0]{\@secondoftwo}%
\providecommand \bibfield  [0]{\@secondoftwo}%
\providecommand \translation [1]{[#1]}%
\providecommand \BibitemOpen [0]{}%
\providecommand \bibitemStop [0]{}%
\providecommand \bibitemNoStop [0]{.\EOS\space}%
\providecommand \EOS [0]{\spacefactor3000\relax}%
\providecommand \BibitemShut  [1]{\csname bibitem#1\endcsname}%
\let\auto@bib@innerbib\@empty
\bibitem [{\citenamefont {Miron}\ \emph {et~al.}(2011)\citenamefont {Miron},
  \citenamefont {Garello}, \citenamefont {Gaudin}, \citenamefont {Zermatten},
  \citenamefont {Costache}, \citenamefont {Auffret}, \citenamefont {Bandiera},
  \citenamefont {Rodmacq}, \citenamefont {Schuhl},\ and\ \citenamefont
  {Gambardella}}]{miron2011perpendicular}%
  \BibitemOpen
  \bibfield  {author} {\bibinfo {author} {\bibfnamefont {I.~M.}\ \bibnamefont
  {Miron}}, \bibinfo {author} {\bibfnamefont {K.}~\bibnamefont {Garello}},
  \bibinfo {author} {\bibfnamefont {G.}~\bibnamefont {Gaudin}}, \bibinfo
  {author} {\bibfnamefont {P.-J.}\ \bibnamefont {Zermatten}}, \bibinfo {author}
  {\bibfnamefont {M.~V.}\ \bibnamefont {Costache}}, \bibinfo {author}
  {\bibfnamefont {S.}~\bibnamefont {Auffret}}, \bibinfo {author} {\bibfnamefont
  {S.}~\bibnamefont {Bandiera}}, \bibinfo {author} {\bibfnamefont
  {B.}~\bibnamefont {Rodmacq}}, \bibinfo {author} {\bibfnamefont
  {A.}~\bibnamefont {Schuhl}}, \ and\ \bibinfo {author} {\bibfnamefont
  {P.}~\bibnamefont {Gambardella}},\ }\href@noop {} {\bibfield  {journal}
  {\bibinfo  {journal} {Nature}\ }\textbf {\bibinfo {volume} {476}},\ \bibinfo
  {pages} {189} (\bibinfo {year} {2011})}\BibitemShut {NoStop}%
\bibitem [{\citenamefont {Liu}\ \emph {et~al.}(2012{\natexlab{a}})\citenamefont
  {Liu}, \citenamefont {Pai}, \citenamefont {Li}, \citenamefont {Tseng},
  \citenamefont {Ralph},\ and\ \citenamefont {Buhrman}}]{Liu2012}%
  \BibitemOpen
  \bibfield  {author} {\bibinfo {author} {\bibfnamefont {L.}~\bibnamefont
  {Liu}}, \bibinfo {author} {\bibfnamefont {C.-F.}\ \bibnamefont {Pai}},
  \bibinfo {author} {\bibfnamefont {Y.}~\bibnamefont {Li}}, \bibinfo {author}
  {\bibfnamefont {H.}~\bibnamefont {Tseng}}, \bibinfo {author} {\bibfnamefont
  {D.}~\bibnamefont {Ralph}}, \ and\ \bibinfo {author} {\bibfnamefont
  {R.}~\bibnamefont {Buhrman}},\ }\href@noop {} {\bibfield  {journal} {\bibinfo
   {journal} {Science}\ }\textbf {\bibinfo {volume} {336}},\ \bibinfo {pages}
  {555} (\bibinfo {year} {2012}{\natexlab{a}})}\BibitemShut {NoStop}%
\bibitem [{\citenamefont {Kobs}\ \emph {et~al.}(2011)\citenamefont {Kobs},
  \citenamefont {He{\ss}e}, \citenamefont {Kreuzpaintner}, \citenamefont
  {Winkler}, \citenamefont {Lott}, \citenamefont {Weinberger}, \citenamefont
  {Schreyer},\ and\ \citenamefont {Oepen}}]{Kobs2011}%
  \BibitemOpen
  \bibfield  {author} {\bibinfo {author} {\bibfnamefont {A.}~\bibnamefont
  {Kobs}}, \bibinfo {author} {\bibfnamefont {S.}~\bibnamefont {He{\ss}e}},
  \bibinfo {author} {\bibfnamefont {W.}~\bibnamefont {Kreuzpaintner}}, \bibinfo
  {author} {\bibfnamefont {G.}~\bibnamefont {Winkler}}, \bibinfo {author}
  {\bibfnamefont {D.}~\bibnamefont {Lott}}, \bibinfo {author} {\bibfnamefont
  {P.}~\bibnamefont {Weinberger}}, \bibinfo {author} {\bibfnamefont
  {A.}~\bibnamefont {Schreyer}}, \ and\ \bibinfo {author} {\bibfnamefont
  {H.}~\bibnamefont {Oepen}},\ }\href@noop {} {\bibfield  {journal} {\bibinfo
  {journal} {Physical review letters}\ }\textbf {\bibinfo {volume} {106}},\
  \bibinfo {pages} {217207} (\bibinfo {year} {2011})}\BibitemShut {NoStop}%
\bibitem [{\citenamefont {Nakayama}\ \emph {et~al.}(2013)\citenamefont
  {Nakayama}, \citenamefont {Althammer}, \citenamefont {Chen}, \citenamefont
  {Uchida}, \citenamefont {Kajiwara}, \citenamefont {Kikuchi}, \citenamefont
  {Ohtani}, \citenamefont {Gepr{\"a}gs}, \citenamefont {Opel}, \citenamefont
  {Takahashi} \emph {et~al.}}]{Nakayama2013}%
  \BibitemOpen
  \bibfield  {author} {\bibinfo {author} {\bibfnamefont {H.}~\bibnamefont
  {Nakayama}}, \bibinfo {author} {\bibfnamefont {M.}~\bibnamefont {Althammer}},
  \bibinfo {author} {\bibfnamefont {Y.-T.}\ \bibnamefont {Chen}}, \bibinfo
  {author} {\bibfnamefont {K.}~\bibnamefont {Uchida}}, \bibinfo {author}
  {\bibfnamefont {Y.}~\bibnamefont {Kajiwara}}, \bibinfo {author}
  {\bibfnamefont {D.}~\bibnamefont {Kikuchi}}, \bibinfo {author} {\bibfnamefont
  {T.}~\bibnamefont {Ohtani}}, \bibinfo {author} {\bibfnamefont
  {S.}~\bibnamefont {Gepr{\"a}gs}}, \bibinfo {author} {\bibfnamefont
  {M.}~\bibnamefont {Opel}}, \bibinfo {author} {\bibfnamefont {S.}~\bibnamefont
  {Takahashi}},  \emph {et~al.},\ }\href@noop {} {\bibfield  {journal}
  {\bibinfo  {journal} {Physical review letters}\ }\textbf {\bibinfo {volume}
  {110}},\ \bibinfo {pages} {206601} (\bibinfo {year} {2013})}\BibitemShut
  {NoStop}%
\bibitem [{\citenamefont {Avci}\ \emph {et~al.}(2015)\citenamefont {Avci},
  \citenamefont {Garello}, \citenamefont {Ghosh}, \citenamefont {Gabureac},
  \citenamefont {Alvarado},\ and\ \citenamefont {Gambardella}}]{Avci2015}%
  \BibitemOpen
  \bibfield  {author} {\bibinfo {author} {\bibfnamefont {C.~O.}\ \bibnamefont
  {Avci}}, \bibinfo {author} {\bibfnamefont {K.}~\bibnamefont {Garello}},
  \bibinfo {author} {\bibfnamefont {A.}~\bibnamefont {Ghosh}}, \bibinfo
  {author} {\bibfnamefont {M.}~\bibnamefont {Gabureac}}, \bibinfo {author}
  {\bibfnamefont {S.~F.}\ \bibnamefont {Alvarado}}, \ and\ \bibinfo {author}
  {\bibfnamefont {P.}~\bibnamefont {Gambardella}},\ }\href@noop {} {\bibfield
  {journal} {\bibinfo  {journal} {Nature Physics}\ }\textbf {\bibinfo {volume}
  {11}},\ \bibinfo {pages} {570} (\bibinfo {year} {2015})}\BibitemShut
  {NoStop}%
\bibitem [{\citenamefont {Olejn{\'\i}k}\ \emph {et~al.}(2015)\citenamefont
  {Olejn{\'\i}k}, \citenamefont {Nov{\'a}k}, \citenamefont {Wunderlich},\ and\
  \citenamefont {Jungwirth}}]{olejnik2015electrical}%
  \BibitemOpen
  \bibfield  {author} {\bibinfo {author} {\bibfnamefont {K.}~\bibnamefont
  {Olejn{\'\i}k}}, \bibinfo {author} {\bibfnamefont {V.}~\bibnamefont
  {Nov{\'a}k}}, \bibinfo {author} {\bibfnamefont {J.}~\bibnamefont
  {Wunderlich}}, \ and\ \bibinfo {author} {\bibfnamefont {T.}~\bibnamefont
  {Jungwirth}},\ }\href@noop {} {\bibfield  {journal} {\bibinfo  {journal}
  {Physical Review B}\ }\textbf {\bibinfo {volume} {91}},\ \bibinfo {pages}
  {180402} (\bibinfo {year} {2015})}\BibitemShut {NoStop}%
\bibitem [{\citenamefont {Yasuda}\ \emph {et~al.}(2016)\citenamefont {Yasuda},
  \citenamefont {Tsukazaki}, \citenamefont {Yoshimi}, \citenamefont
  {Takahashi}, \citenamefont {Kawasaki},\ and\ \citenamefont
  {Tokura}}]{yasuda2016large}%
  \BibitemOpen
  \bibfield  {author} {\bibinfo {author} {\bibfnamefont {K.}~\bibnamefont
  {Yasuda}}, \bibinfo {author} {\bibfnamefont {A.}~\bibnamefont {Tsukazaki}},
  \bibinfo {author} {\bibfnamefont {R.}~\bibnamefont {Yoshimi}}, \bibinfo
  {author} {\bibfnamefont {K.}~\bibnamefont {Takahashi}}, \bibinfo {author}
  {\bibfnamefont {M.}~\bibnamefont {Kawasaki}}, \ and\ \bibinfo {author}
  {\bibfnamefont {Y.}~\bibnamefont {Tokura}},\ }\href@noop {} {\bibfield
  {journal} {\bibinfo  {journal} {Physical Review Letters}\ }\textbf {\bibinfo
  {volume} {117}},\ \bibinfo {pages} {127202} (\bibinfo {year}
  {2016})}\BibitemShut {NoStop}%
\bibitem [{\citenamefont {Zhang}\ and\ \citenamefont
  {Vignale}(2016)}]{Zhang2016a}%
  \BibitemOpen
  \bibfield  {author} {\bibinfo {author} {\bibfnamefont {S.~S.-L.}\
  \bibnamefont {Zhang}}\ and\ \bibinfo {author} {\bibfnamefont
  {G.}~\bibnamefont {Vignale}},\ }\href@noop {} {\bibfield  {journal} {\bibinfo
   {journal} {Physical Review B}\ }\textbf {\bibinfo {volume} {94}},\ \bibinfo
  {pages} {140411} (\bibinfo {year} {2016})}\BibitemShut {NoStop}%
\bibitem [{\citenamefont {Kobs}\ and\ \citenamefont {Oepen}(2016)}]{Kobs2016}%
  \BibitemOpen
  \bibfield  {author} {\bibinfo {author} {\bibfnamefont {A.}~\bibnamefont
  {Kobs}}\ and\ \bibinfo {author} {\bibfnamefont {H.~P.}\ \bibnamefont
  {Oepen}},\ }\href@noop {} {\bibfield  {journal} {\bibinfo  {journal}
  {Physical Review B}\ }\textbf {\bibinfo {volume} {93}},\ \bibinfo {pages}
  {014426} (\bibinfo {year} {2016})}\BibitemShut {NoStop}%
\bibitem [{\citenamefont {Nernst}(1887)}]{nernst1887electromotorischen}%
  \BibitemOpen
  \bibfield  {author} {\bibinfo {author} {\bibfnamefont {W.}~\bibnamefont
  {Nernst}},\ }\href@noop {} {\bibfield  {journal} {\bibinfo  {journal}
  {Annalen der Physik}\ }\textbf {\bibinfo {volume} {267}},\ \bibinfo {pages}
  {760} (\bibinfo {year} {1887})}\BibitemShut {NoStop}%
\bibitem [{\citenamefont {Avci}\ \emph {et~al.}(2014)\citenamefont {Avci},
  \citenamefont {Garello}, \citenamefont {Gabureac}, \citenamefont {Ghosh},
  \citenamefont {Fuhrer}, \citenamefont {Alvarado},\ and\ \citenamefont
  {Gambardella}}]{Avci2014a}%
  \BibitemOpen
  \bibfield  {author} {\bibinfo {author} {\bibfnamefont {C.~O.}\ \bibnamefont
  {Avci}}, \bibinfo {author} {\bibfnamefont {K.}~\bibnamefont {Garello}},
  \bibinfo {author} {\bibfnamefont {M.}~\bibnamefont {Gabureac}}, \bibinfo
  {author} {\bibfnamefont {A.}~\bibnamefont {Ghosh}}, \bibinfo {author}
  {\bibfnamefont {A.}~\bibnamefont {Fuhrer}}, \bibinfo {author} {\bibfnamefont
  {S.~F.}\ \bibnamefont {Alvarado}}, \ and\ \bibinfo {author} {\bibfnamefont
  {P.}~\bibnamefont {Gambardella}},\ }\href@noop {} {\bibfield  {journal}
  {\bibinfo  {journal} {Physical Review B}\ }\textbf {\bibinfo {volume} {90}},\
  \bibinfo {pages} {224427} (\bibinfo {year} {2014})}\BibitemShut {NoStop}%
\bibitem [{\citenamefont {Agustsson}\ \emph {et~al.}(2008)\citenamefont
  {Agustsson}, \citenamefont {Arnalds}, \citenamefont {Ingason}, \citenamefont
  {Gylfason}, \citenamefont {Johnsen}, \citenamefont {Olafsson},\ and\
  \citenamefont {Gudmundsson}}]{agustsson2008electrical}%
  \BibitemOpen
  \bibfield  {author} {\bibinfo {author} {\bibfnamefont {J.}~\bibnamefont
  {Agustsson}}, \bibinfo {author} {\bibfnamefont {U.}~\bibnamefont {Arnalds}},
  \bibinfo {author} {\bibfnamefont {A.}~\bibnamefont {Ingason}}, \bibinfo
  {author} {\bibfnamefont {K.~B.}\ \bibnamefont {Gylfason}}, \bibinfo {author}
  {\bibfnamefont {K.}~\bibnamefont {Johnsen}}, \bibinfo {author} {\bibfnamefont
  {S.}~\bibnamefont {Olafsson}}, \ and\ \bibinfo {author} {\bibfnamefont
  {J.~T.}\ \bibnamefont {Gudmundsson}},\ }\href@noop {} {\bibfield  {journal}
  {\bibinfo  {journal} {Journal of Physics: Conference Series}\ }\textbf
  {\bibinfo {volume} {100}},\ \bibinfo {pages} {082006} (\bibinfo {year}
  {2008})}\BibitemShut {NoStop}%
\bibitem [{\citenamefont {Carcia}(1988)}]{carcia1988perpendicular}%
  \BibitemOpen
  \bibfield  {author} {\bibinfo {author} {\bibfnamefont {P.}~\bibnamefont
  {Carcia}},\ }\href@noop {} {\bibfield  {journal} {\bibinfo  {journal}
  {Journal of applied physics}\ }\textbf {\bibinfo {volume} {63}},\ \bibinfo
  {pages} {5066} (\bibinfo {year} {1988})}\BibitemShut {NoStop}%
\bibitem [{\citenamefont {Zhang}\ \emph {et~al.}(2013)\citenamefont {Zhang},
  \citenamefont {Vlaminck}, \citenamefont {Pearson}, \citenamefont {Divan},
  \citenamefont {Bader},\ and\ \citenamefont
  {Hoffmann}}]{zhang2013determination}%
  \BibitemOpen
  \bibfield  {author} {\bibinfo {author} {\bibfnamefont {W.}~\bibnamefont
  {Zhang}}, \bibinfo {author} {\bibfnamefont {V.}~\bibnamefont {Vlaminck}},
  \bibinfo {author} {\bibfnamefont {J.~E.}\ \bibnamefont {Pearson}}, \bibinfo
  {author} {\bibfnamefont {R.}~\bibnamefont {Divan}}, \bibinfo {author}
  {\bibfnamefont {S.~D.}\ \bibnamefont {Bader}}, \ and\ \bibinfo {author}
  {\bibfnamefont {A.}~\bibnamefont {Hoffmann}},\ }\href@noop {} {\bibfield
  {journal} {\bibinfo  {journal} {Applied physics letters}\ }\textbf {\bibinfo
  {volume} {103}},\ \bibinfo {pages} {242414} (\bibinfo {year}
  {2013})}\BibitemShut {NoStop}%
\bibitem [{\citenamefont {Piraux}\ \emph {et~al.}(1998)\citenamefont {Piraux},
  \citenamefont {Dubois}, \citenamefont {Fert},\ and\ \citenamefont
  {Belliard}}]{piraux1998temperature}%
  \BibitemOpen
  \bibfield  {author} {\bibinfo {author} {\bibfnamefont {L.}~\bibnamefont
  {Piraux}}, \bibinfo {author} {\bibfnamefont {S.}~\bibnamefont {Dubois}},
  \bibinfo {author} {\bibfnamefont {A.}~\bibnamefont {Fert}}, \ and\ \bibinfo
  {author} {\bibfnamefont {L.}~\bibnamefont {Belliard}},\ }\href@noop {}
  {\bibfield  {journal} {\bibinfo  {journal} {The European Physical Journal
  B-Condensed Matter and Complex Systems}\ }\textbf {\bibinfo {volume} {4}},\
  \bibinfo {pages} {413} (\bibinfo {year} {1998})}\BibitemShut {NoStop}%
\bibitem [{\citenamefont {Kim}\ \emph {et~al.}(2016{\natexlab{a}})\citenamefont
  {Kim}, \citenamefont {Moriyama}, \citenamefont {Koyama}, \citenamefont
  {Chiba}, \citenamefont {Lee}, \citenamefont {Lee}, \citenamefont {Lee},
  \citenamefont {Lee},\ and\ \citenamefont {Ono}}]{kim2016current}%
  \BibitemOpen
  \bibfield  {author} {\bibinfo {author} {\bibfnamefont {K.}~\bibnamefont
  {Kim}}, \bibinfo {author} {\bibfnamefont {T.}~\bibnamefont {Moriyama}},
  \bibinfo {author} {\bibfnamefont {T.}~\bibnamefont {Koyama}}, \bibinfo
  {author} {\bibfnamefont {D.}~\bibnamefont {Chiba}}, \bibinfo {author}
  {\bibfnamefont {S.}~\bibnamefont {Lee}}, \bibinfo {author} {\bibfnamefont
  {S.}~\bibnamefont {Lee}}, \bibinfo {author} {\bibfnamefont {K.}~\bibnamefont
  {Lee}}, \bibinfo {author} {\bibfnamefont {H.}~\bibnamefont {Lee}}, \ and\
  \bibinfo {author} {\bibfnamefont {T.}~\bibnamefont {Ono}},\ }\href@noop {}
  {\bibfield  {journal} {\bibinfo  {journal} {arXiv arXiv:1603.08746}\ }
  (\bibinfo {year} {2016}{\natexlab{a}})}\BibitemShut {NoStop}%
\bibitem [{\citenamefont {Isasa}\ \emph {et~al.}(2015)\citenamefont {Isasa},
  \citenamefont {Villamor}, \citenamefont {Hueso}, \citenamefont {Gradhand},\
  and\ \citenamefont {Casanova}}]{isasa2015temperature}%
  \BibitemOpen
  \bibfield  {author} {\bibinfo {author} {\bibfnamefont {M.}~\bibnamefont
  {Isasa}}, \bibinfo {author} {\bibfnamefont {E.}~\bibnamefont {Villamor}},
  \bibinfo {author} {\bibfnamefont {L.~E.}\ \bibnamefont {Hueso}}, \bibinfo
  {author} {\bibfnamefont {M.}~\bibnamefont {Gradhand}}, \ and\ \bibinfo
  {author} {\bibfnamefont {F.}~\bibnamefont {Casanova}},\ }\href@noop {}
  {\bibfield  {journal} {\bibinfo  {journal} {Physical Review B}\ }\textbf
  {\bibinfo {volume} {91}},\ \bibinfo {pages} {024402} (\bibinfo {year}
  {2015})}\BibitemShut {NoStop}%
\bibitem [{\citenamefont {Ueda}\ \emph {et~al.}(2012)\citenamefont {Ueda},
  \citenamefont {Koyama}, \citenamefont {Hiramatsu}, \citenamefont {Chiba},
  \citenamefont {Fukami}, \citenamefont {Tanigawa}, \citenamefont {Suzuki},
  \citenamefont {Ohshima}, \citenamefont {Ishiwata}, \citenamefont {Nakatani}
  \emph {et~al.}}]{ueda2012temperature}%
  \BibitemOpen
  \bibfield  {author} {\bibinfo {author} {\bibfnamefont {K.}~\bibnamefont
  {Ueda}}, \bibinfo {author} {\bibfnamefont {T.}~\bibnamefont {Koyama}},
  \bibinfo {author} {\bibfnamefont {R.}~\bibnamefont {Hiramatsu}}, \bibinfo
  {author} {\bibfnamefont {D.}~\bibnamefont {Chiba}}, \bibinfo {author}
  {\bibfnamefont {S.}~\bibnamefont {Fukami}}, \bibinfo {author} {\bibfnamefont
  {H.}~\bibnamefont {Tanigawa}}, \bibinfo {author} {\bibfnamefont
  {T.}~\bibnamefont {Suzuki}}, \bibinfo {author} {\bibfnamefont
  {N.}~\bibnamefont {Ohshima}}, \bibinfo {author} {\bibfnamefont
  {N.}~\bibnamefont {Ishiwata}}, \bibinfo {author} {\bibfnamefont
  {Y.}~\bibnamefont {Nakatani}},  \emph {et~al.},\ }\href@noop {} {\bibfield
  {journal} {\bibinfo  {journal} {Applied Physics Letters}\ }\textbf {\bibinfo
  {volume} {100}},\ \bibinfo {pages} {202407} (\bibinfo {year}
  {2012})}\BibitemShut {NoStop}%
\bibitem [{\citenamefont {Kim}\ \emph {et~al.}(2016{\natexlab{b}})\citenamefont
  {Kim}, \citenamefont {Sheng}, \citenamefont {Takahashi}, \citenamefont
  {Mitani},\ and\ \citenamefont {Hayashi}}]{Kim2015}%
  \BibitemOpen
  \bibfield  {author} {\bibinfo {author} {\bibfnamefont {J.}~\bibnamefont
  {Kim}}, \bibinfo {author} {\bibfnamefont {P.}~\bibnamefont {Sheng}}, \bibinfo
  {author} {\bibfnamefont {S.}~\bibnamefont {Takahashi}}, \bibinfo {author}
  {\bibfnamefont {S.}~\bibnamefont {Mitani}}, \ and\ \bibinfo {author}
  {\bibfnamefont {M.}~\bibnamefont {Hayashi}},\ }\href@noop {} {\bibfield
  {journal} {\bibinfo  {journal} {Physical review letters}\ }\textbf {\bibinfo
  {volume} {116}},\ \bibinfo {pages} {097201} (\bibinfo {year}
  {2016}{\natexlab{b}})}\BibitemShut {NoStop}%
\bibitem [{\citenamefont {Liu}\ \emph {et~al.}(2012{\natexlab{b}})\citenamefont
  {Liu}, \citenamefont {Lee}, \citenamefont {Gudmundsen}, \citenamefont
  {Ralph},\ and\ \citenamefont {Buhrman}}]{Liu2012a}%
  \BibitemOpen
  \bibfield  {author} {\bibinfo {author} {\bibfnamefont {L.}~\bibnamefont
  {Liu}}, \bibinfo {author} {\bibfnamefont {O.~J.}\ \bibnamefont {Lee}},
  \bibinfo {author} {\bibfnamefont {T.~J.}\ \bibnamefont {Gudmundsen}},
  \bibinfo {author} {\bibfnamefont {D.~C.}\ \bibnamefont {Ralph}}, \ and\
  \bibinfo {author} {\bibfnamefont {R.~a.}\ \bibnamefont {Buhrman}},\ }\href
  {\doibase 10.1103/PhysRevLett.109.096602} {\bibfield  {journal} {\bibinfo
  {journal} {Phys. Rev. Lett.}\ }\textbf {\bibinfo {volume} {109}},\ \bibinfo
  {pages} {1} (\bibinfo {year} {2012}{\natexlab{b}})}\BibitemShut {NoStop}%
\bibitem [{\citenamefont {Zhang}\ \emph {et~al.}(2015)\citenamefont {Zhang},
  \citenamefont {Han}, \citenamefont {Jiang}, \citenamefont {Yang},\ and\
  \citenamefont {Parkin}}]{zhang2015role}%
  \BibitemOpen
  \bibfield  {author} {\bibinfo {author} {\bibfnamefont {W.}~\bibnamefont
  {Zhang}}, \bibinfo {author} {\bibfnamefont {W.}~\bibnamefont {Han}}, \bibinfo
  {author} {\bibfnamefont {X.}~\bibnamefont {Jiang}}, \bibinfo {author}
  {\bibfnamefont {S.-H.}\ \bibnamefont {Yang}}, \ and\ \bibinfo {author}
  {\bibfnamefont {S.~S.}\ \bibnamefont {Parkin}},\ }\href@noop {} {\bibfield
  {journal} {\bibinfo  {journal} {Nature Physics}\ }\textbf {\bibinfo {volume}
  {11}},\ \bibinfo {pages} {496} (\bibinfo {year} {2015})}\BibitemShut
  {NoStop}%
\bibitem [{\citenamefont {Amin}\ and\ \citenamefont
  {Stiles}(2016)}]{amin2016spin}%
  \BibitemOpen
  \bibfield  {author} {\bibinfo {author} {\bibfnamefont {V.~P.}\ \bibnamefont
  {Amin}}\ and\ \bibinfo {author} {\bibfnamefont {M.~D.}\ \bibnamefont
  {Stiles}},\ }\href@noop {} {\bibfield  {journal} {\bibinfo  {journal}
  {Physical Review B}\ }\textbf {\bibinfo {volume} {94}},\ \bibinfo {pages}
  {104419} (\bibinfo {year} {2016})}\BibitemShut {NoStop}%
\bibitem [{\citenamefont {Wang}\ \emph {et~al.}(2016)\citenamefont {Wang},
  \citenamefont {Wesselink}, \citenamefont {Liu}, \citenamefont {Yuan},
  \citenamefont {Xia},\ and\ \citenamefont {Kelly}}]{wang2016giant}%
  \BibitemOpen
  \bibfield  {author} {\bibinfo {author} {\bibfnamefont {L.}~\bibnamefont
  {Wang}}, \bibinfo {author} {\bibfnamefont {R.}~\bibnamefont {Wesselink}},
  \bibinfo {author} {\bibfnamefont {Y.}~\bibnamefont {Liu}}, \bibinfo {author}
  {\bibfnamefont {Z.}~\bibnamefont {Yuan}}, \bibinfo {author} {\bibfnamefont
  {K.}~\bibnamefont {Xia}}, \ and\ \bibinfo {author} {\bibfnamefont {P.~J.}\
  \bibnamefont {Kelly}},\ }\href@noop {} {\bibfield  {journal} {\bibinfo
  {journal} {Physical review letters}\ }\textbf {\bibinfo {volume} {116}},\
  \bibinfo {pages} {196602} (\bibinfo {year} {2016})}\BibitemShut {NoStop}%
\end{thebibliography}%

\end{document}